\documentclass[referee]{aa}
\usepackage{graphicx}
\begin{document}
\title{An ISO--SWS survey of molecular hydrogen in Starburst and Seyfert
Galaxies
\thanks{Based on observations with ISO, an ESA project with instruments funded
by ESA Member States (especially the PI countries: France, Germany, the
Netherlands and the United Kingdom) and with the participation of ISAS and
NASA.} }
\author{D. Rigopoulou\inst{1}
\and D. Kunze\inst{1}
\and D. Lutz\inst{1}
\and R. Genzel\inst{1}
\and A.F.M. Moorwood \inst{2}}
\institute{Max-Planck-Institut f\"ur extraterrestrische Physik,
Postfach 1312, 85741 Garching bei M\"unchen, Germany
\and European Southern Observatory, Karl-Schwarzschild-Str. 2, 85748 Garching 
bei M\"unchen, Germany
}
\offprints{D. Rigopoulou (MPE, Germany), \email{dar@mpe.mpg.de}}
\date{Received  / Accepted }
\titlerunning{H$_{2}$ observations in Starbursts and Seyferts}
\abstract{ We present results from a survey of molecular hydrogen 
emission from a 
sample of Starburst and Seyfert galaxies carried out with the Infrared
Space Observatory (ISO). 
Pure rotational H$_{2}$ emission has been detected
in a number of extragalactic objects and a 
variety of environments. A number of transitions from S(7) to S(0) are detected
in both Starbursts and Seyferts.
Using excitation diagrams we derive temperatures and masses of the ``warm'' 
molecular hydrogen. 
We find that the temperature of the ``warm'' gas is similar 
in Starbursts and Seyferts (those Seyferts for which we have firm detections
of the S(0) line) with a value of around T$\sim$150 K. 
This ``warm'' gas accounts for as much as 10\% of the 
total galactic mass (as probed by CO molecular observations) in Starbursts.
The fraction of ``warm'' gas is overall higher in Seyferts, 
ranging between 2--35\%.
We then investigate the origin of the warm H$_{2}$ emission. 
Comparison with published theoretical models and Galactic templates 
implies that
although emission from photodissociation regions (PDR) alone could 
explain the emission from Starbursts 
and Seyferts,
most likely a combination of PDR, shock emission and gas heated by X-rays
(mostly for the Seyferts)  
is responsible for H$_{2}$ excitation in extragalactic environments. 
Finally, we find that although PAH and H$_{2}$ line emission correlate
well in Starbursts and the large scale emission in AGN, H$_{2}$ emission
is much stronger compared to PAH emission in cases where a ``pure'' 
AGN dominates the energy output. 
}
\maketitle

\section{Introduction}

Using the Short Wavelength Spectrometer (SWS; De Graauw et al. 1996) on board 
the Infrared Space Observatory (ISO; Kessler et al. 1996), 
we have performed a survey of molecular hydrogen emission from 
active galaxies displaying a wide range in nuclear activity including
pure bona-fide Starbursts, Seyfert 2s (some of them with starburst components)
and pure Seyfert 1 galaxies.
Prior to the ISO mission, extragalactic H$_{2}$ emission
had only been detected in the ro-vibrational 
lines around 2.1 $\mu$m. Indeed, H$_{2}$ ro-vibrational emission has been
detected in galactic sources (e.g. Usuda et al., 1996), 
Starburst galaxies (e.g. Joseph, Wright and Wade 1984), 
Seyferts (e.g. Moorwood and Oliva 1988, Fischer et al. 1987 ) 
and bright spirals 
(e.g. Puxley, Hawarden and Mountain 1988).

The ro-vibrational lines typically trace H$_{2}$ gas of
masses around $\sim$10$^{4-5}$ M$_{\odot}$ and temperatures $\sim$2000 K. This
gas can be excited either by collisions (thermal) or by absorption of 
ultraviolet (UV)
photons in the Lyman and Werner electronic bands (912-1108 \.A), followed by a
de-excitation cascade to the ground state (fluorescence). 
However, gas at these temperatures is a very small fraction 
(as small as 10$^{-6}$) of the total amount of H$_{2}$ gas
(e.g. Van der Werf et al. 1993). 
Since the ro-vibrational lines tend to get faint at lower temperatures,
most of our knowledge about the H$_{2}$ content of galaxies comes from CO
observations assuming a CO/H$_{2}$ conversion factor derived from galactic 
molecular
cloud observations. ISO gave the unique opportunity to observe intermediate
temperature gas, ie ``warm'' H$_{2}$, directly in pure rotational lines. Since 
transitions with $\Delta$J= $\pm$ 1 are strictly forbidden for the H$_{2}$
molecule, the rotational ladder consists
only of an ortho (J odd) and a para (J even) series of quadrupole transitions.

ISO and in particular SWS, offered the unique opportunity to detect
pure rotational H$_{2}$ emission in a number of
galactic and  extragalactic sources, thus studying the amount of
moderately warm gas in these sources. The spectral range of SWS
provides full coverage of a number of transitions (for most galaxies
we have observed from the S(0) to S(7) transitions)
while its spectral resolution is well matched to the typical velocity
dispersions of galaxies.  Among the first detections of pure
rotational H$_{2}$ emission in galaxies were the detections in NGC
3256 (Rigopoulou et al. 1996), NGC 6946 (Valentijn et al.
1996), NGC 891 (Valentijn \& Van der Werf 1999).
However, no study of pure rotational H$_{2}$
emission for a large number of galaxies has so far appeared. 

Here, we present an
inventory of H$_{2}$ emission lines from a number of Starburst and
AGN. Our survey includes 12 and 9 Starburst\footnote {The two components
of NGC 3690 have been observed separately NGC 3690A and NGC 3690BC. In the
analysis we treat them as two separate systems.} and Seyfert 
galaxies, respectively. Temperatures of the warm molecular gas are deduced from
excitation diagrams whereas the masses of the warm molecular gas are
compared to the total gas content of the galaxies as estimated from 
molecular CO observations.
The H$_{2}$ excitation mechanism is investigated next. 
The observations are compared to the predictions of published models
both for PDR, shocked emission and X-ray irradiated gas, as well as to
Galactic templates. Finally we examine possible correlations between
PAH and H$_{2}$ emission. 

\section{Observations and data reduction}

The data presented here are part of the MPE ISO Guaranteed time
project "Bright Galactic Nuclei". The observations were carried out
throughout the ISO mission and were all taken in the standard AOT
SWS02 mode, i.e. grating line profile scan. 
The SWS has three different apertures with dimensions 
(projected on the sky) ap1: 14''$\times$20'',
ap2: 14''$\times$27'' and ap3: 14''$\times$20''.
The following molecular 
transitions (and the corresponding apertures) were observed although 
not in all sources:
(1--0)Q(3) (ap1), (0--0)S(0) (ap3), 
(0--0)S(1) (ap2), (0--0)S(2) (ap1), (0--0)S(3) (ap3), 
(0--0)S(5) (ap2), and 
(0--0)S(7) (ap2). 
The sample targets are presented in Tables 1 and 2 for the Starbursts and
Seyferts, respectively.
Galaxy names, distances (assuming H$_{\small 0}$ = 75
kms$^{-1}$ Mpc$^{-1}$), extinction towards the ionized medium, and 
comments are also listed. 
We note that the SWS aperture sizes are most likely larger than 
the presumed 
sizes of the circumnuclear regions for a large fraction of the present samples
of Starburst and Seyfert galaxies.

\begin{table*}
\caption[]{The Starburst sample galaxies}
\begin{flushleft}
\begin{tabular}{cccccccc}
 & & & & & & & \\ \hline 
\hline 
\\ Galaxy Name&D&A$_{\small V} ^{1}$&Spec. type&comments\\ 
& Mpc&[mag]& & \\ \\ \hline 
NGC 253&3.5&30&HII, SB&Sc, edge-on \\ 
IC 342&3.6&12&HII, SB&Scd, face-on\\
II Zw 40&7&3$^{2}$ &HII&BCD\\ 
M 82$^{3}$&3.3 &10&HII, SB&Irr \\ 
NGC 3256$^{a}$&36.6&35 &SB&colliding pair\\ 
NGC 3690A$/$BC$^{3}$&40&20&SB& pec. merger \\ 
NGC 4038$/$39$^{3,b}$&21.0&80&SB&interacting \\ 
NGC 4945$^{c}$&4&20 &SB, Sy2&Sc edge-on\\ 
NGC 5236 (M 83)&5.0&5&HII, SB&SAB(s), face-on\\
NGC 5253&4.1&14&HII, SB&Im pec\\ 
NGC 6946$^{d}$&5.7&1.8$^{4}$&HII&S(AB)cd,face-on\\ 
NGC 7552&21.2&5&HII, LINER&SA(c), face--on \\ 
& & & & \\
\hline
\end{tabular}\\
$^{1}$: A$_{\small V}$ from Genzel et al. (1998)\\ 
$^{2}$: A$_{\small V}$ (screen) from Baldwin, Spinrad and Terlevich (1982)\\
$^{3}$: The ISO pointings on these galaxies are as follows (J2000): 
M 82:00 39 43.38, 69 40 44.4, NGC 3690A: 11 45 54.24, 58 33 46.5, NGC 3690BC:
11 45 54.06, 58 33 45.6, NGC 4038$/$39:00 48 07.66,  -18 53 04.1. The ISO
pointings for the remaining galaxies are on the galaxy's nucleus.\\
$^{4}$: A$_{\small V}$ (screen) from Hyman et al. (2000)\\
$^{a}$: Rigopoulou et al. (1996),
$^{b}$: Kunze et al. (1996),
$^{c}$: Spoon et al. (2000),
$^{d}$: Valentijn et al. (1996)\\
\end{flushleft}
\end{table*}

\begin{table*}
\caption[]{\it The Seyfert sample galaxies}
\begin{flushleft}
\begin{tabular}{cccccccc}
 & & & & & & \\
\hline
\hline
Galaxy Name&D&A$_{\small V}$&Spec. type&comments\\
 & Mpc&[mag]& & \\
\hline
NGC 1068&14 &8$^{1}$&Sy2&SAb \\
NGC 1275&70&2$^{2}$ &Sy2& cD, pec. \\
NGC 1365&22.3 &2.5$^{2}$&Sy1.8 &SBb \\
NGC 4151&20 &3$^{1}$&Sy1.5&SABab \\
Cen A&4 &30$^{3}$ &Radio&SO pec. \\
NGC 5506&23 &8$^{3}$ &Sy1.9&Sa pec. \\
Circinus&3&20$^{3}$&Sy2&SAb\\
NGC 7582&20 &18$^{3}$ &Sy2&SBab \\
NGC 7469&66&20$^{1}$&Sy1.2&SABa \\
 & &  & & \\   \hline
\end{tabular}\\
$^{1}$: A$_{\small V}$ mixed case (from Genzel et al. 1998)\\
$^{2}$: NGC 1275: Krabbe et al. (2000); NGC 1365: Kristen et al. (1997)\\
$^{3}$: A$_{\small V}$ screen case (from Genzel et al. 1998)\\
\end{flushleft}
\end{table*}

The data were reduced using the Interactive Analysis
(IA) Package of the SWS team including sophisticated
tools to improve cosmic ray hit removal, dark subtraction,  defringing
and flat-fielding. The flux calibration is accurate to about 25\% (Schaeidt
1996). The calibration files of June 2000 were used.

Figures 1 and 2 show the observed H$_{2}$ molecular lines
in Starburst and Seyfert galaxies\footnote{Previously published H$_{2}$
molecular lines are not shown here. The spectra can be found in:
Rigopoulou et al. (1996) for NGC 3256;  
Kunze et al., (1996) for NGC 4038$/$39; Spoon et al., (2000) for NGC 4945.}.
In almost all cases the 
S(1) and S(5) lines are clearly detected in both the Seyfert and the 
Starburst samples. Tables 3 and 4 give the actual line flux values
for Starbursts and Seyferts, respectively.

From the sample of 13 Starbursts (the two components in NGC 3690 have 
been observed separately) the S(0) line was observed in 10 of them
and detected in 5. The S(0) line was observed in all (9) Seyferts and detected
in 4 of them.

\begin{figure*}
 \includegraphics[width=18cm]{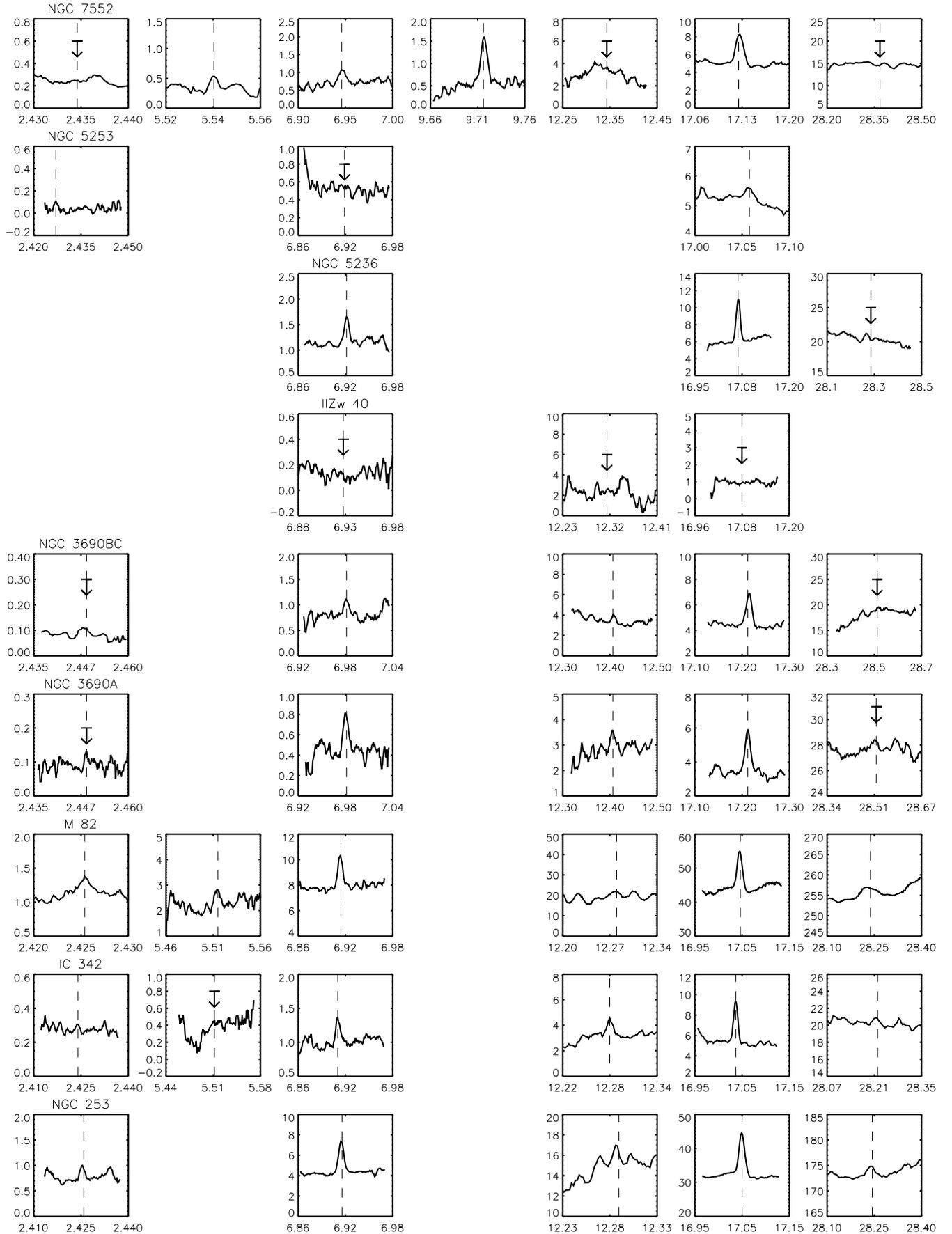}
  \caption{H$_2$ molecular lines from Starbursts: from left to right 
 (1--0)Q(3), (0-0)S(7), (0-0)S(5), (0-0)S(3), (0-0)S(2), (0-0)S(1) and 
 (0-0)S(0). The vertical axis is Flux in Jy, the horizontal axis is wavelength
in $\mu$m.
Note: previously published spectra for NGC 3256, NGC 4038 and 
NGC 4945 and NGC 6946 are not included in this Figure.}
\end{figure*}

\begin{figure*}
\includegraphics[width=18cm]{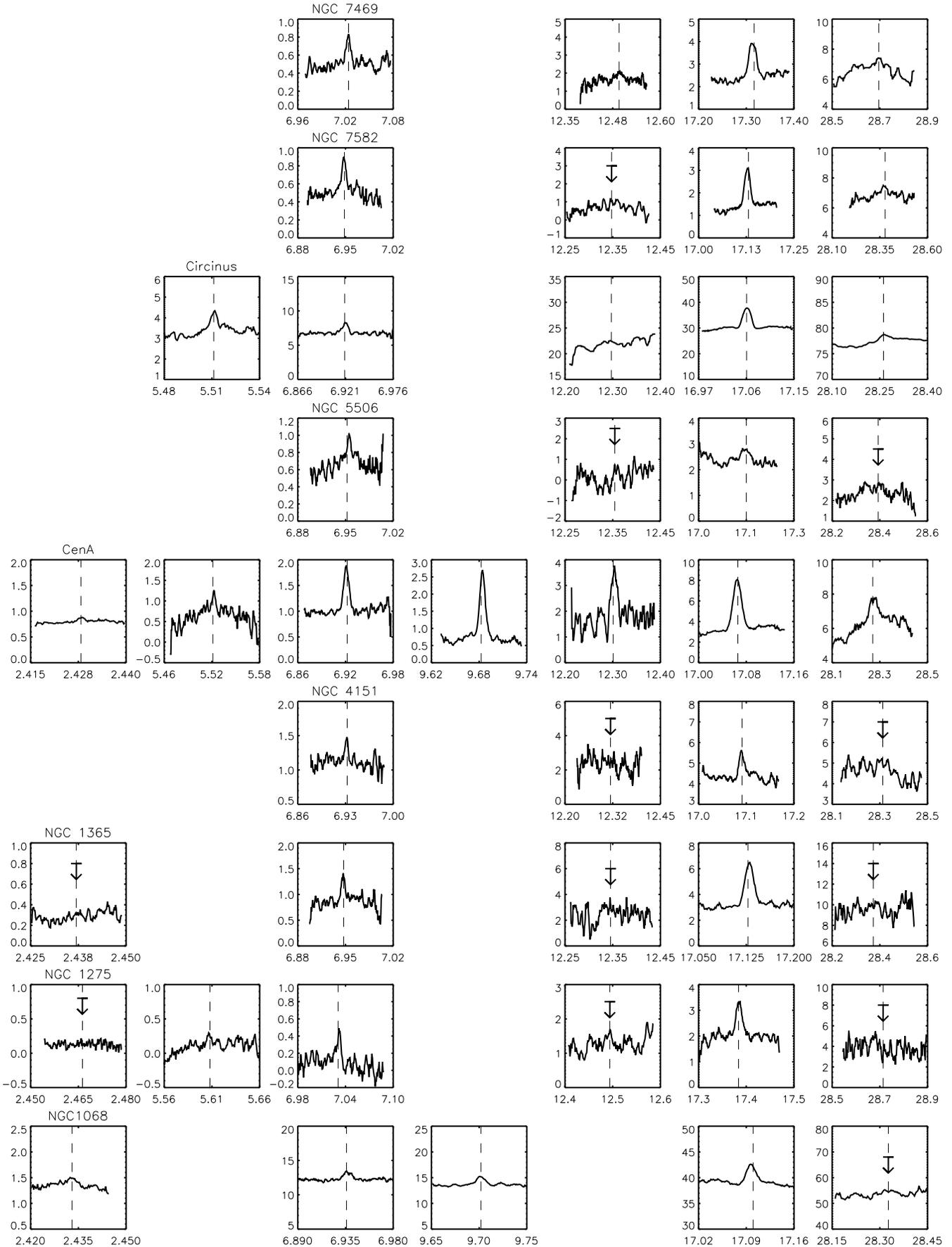}
\caption{H$_2$ molecular lines in Seyferts. The spectral lines from the 
 various transitions and the axes are the same as in Figure 1. }
\end{figure*}

\begin{table*}
\caption[]{\it H$_{2}$ Emission Line Fluxes -- Starbursts}
\begin{flushleft}
\begin{tabular}{cccccccc}
\hline
\hline
Galaxy Name&(1--0)Q(3)&S(7)&S(5)&S(3)&S(2)&S(1)&S(0)\\
$\lambda (\mu m)$ &2.42&5.51&6.91&9.67&12.27&17.03&28.22\\
&  & &$\times$10$^{-20}$ W cm$^{-2}$& & & & \\
\hline
NGC 253&2.80&8.40&11.5&-- &12.0&19.57&2.13\\
IC 342&0.39&$<$0.9&1.60&-- &2.50&4.90&0.80\\
IIZw40&--&--&$<$0.45&--&$<$4.5&$<$0.85&--\\
M 82&2.65&4.80&11.50&--&12.0&15.0&7.8\\
NGC 3256$^{a}$&0.38&--&4.70&--&2.35&11.50&$<$3.90\\
NGC 3690A&$<$0.3&-- &1.97&-- &1.8&3.7&$<$1.21\\
NGC 3690B$/$C&$<$0.4&--&1.46&--&1.92&4.11&$<$0.7\\
NGC 4038$^{b}$&--&--&--&--&1.6&3.95&--\\
NGC 4945$^{c}$&3.2&1.11&15.4& &7.45&15.1&4.82\\
NGC 5236 (M83)&--&--&2.84&--&--&7.29&$<$1.04\\
NGC 5253&0.55&--&$<$0.60&--&--&0.50&--\\
NGC 6946$^{d}$&--&--&--&--&$<$0.93&2.73&0.41\\
NGC 7552&$<$0.86&1.08&2.37&2.84&$<$1.5&5.11&$<$0.52\\
 & & & & &  & \\   \hline
\end{tabular}\\
H$_{2}$ line fluxes from: \\
$^{a}$: Rigopoulou et al. (1996)\\
$^{b}$: Kunze et al. (1996)\\
$^{c}$: Spoon et al. (2000)\\
$^{d}$: Valentijn et al. (1996)\\
\end{flushleft}
\end{table*}

For those  sources where lines were not detected we quote
``good confidence'' upper limits.  As mentioned earlier a
large number of atomic and ionic lines were observed  for each source.
For the purpose of the current paper we focus on the H$_{2}$ emission.
More detailed analysis for some of the sources can be found in other
papers (NGC 4038$/$39: Kunze et al. (1996), NGC 3256: Rigopoulou et
al. 1996, NGC 4151: Alexander et al. (1998), Sturm et al. (1998), Circinus:
Moorwood et al. (1996), NGC 1275: Krabbe et al. (2000); NGC 1068: Lutz et 
al. (2000)).

\begin{table*}
\caption[]{\it H$_{2}$ Emission Line Fluxes -- Seyferts}
\begin{tabular}{cccccccc}
 & & & & & & & \\
\hline
\hline
Galaxy Name&(1--0)Q(3)&S(7)&S(5)&S(3)&S(2)&S(1)&S(0)\\
$\lambda (\mu m)$ &2.42&5.51&6.91&9.67&12.27&17.03&28.22\\
&  & &$\times$10$^{-20}$ W cm$^{-2}$& & & & \\
\hline
NGC 1068$^{a}$&2.296&--&6.403&5.757&--&6.502&$<$1.872\\
NGC 1275&$<$0.8&0.9&1.55&-- &$<$1.0&2.06&$<$2.2\\
NGC 1365&$<$0.85&--&2.01&--&$<$3.13&5.69&$<$1.65\\
NGC 4151&--&--&1.32&--&$<$1.86&1.667&$<$1.26\\
CenA&0.792&3.259&4.535&5.806&5.397&8.635&2.511\\
NGC 5506&--&--&1.508&--&$<$0.859&1.185&$<$0.928\\
Circinus&--&2.72&7.97&--&2.36 &13.94&1.56 \\
NGC 7582&--&--&1.945&--&$<$1.289&3.116&0.764\\
NGC 7469&--&--&1.394&--&2.125&2.955&0.8\\
 & & & & & & & \\  \hline
\end{tabular}\\
H$_{2}$ line fluxes from:\\
$^{a}$: Lutz et al. (2000)\\
\end{table*}

\section{Warm Molecular Hydrogen Emission: Excitation Temperatures and Masses}

Using the pure rotational H$_{2}$ lines we can probe the
physical conditions of the warm molecular hydrogen in our sample of
Starburst and Seyfert galaxies. As we discussed in Section 1., 
the pure rotational lines originate from the warm (T$<$ 2000 K) gas.
In what follows we derive excitation temperatures for
the ``warm'' molecular hydrogen gas as well as masses and compare them to the
total masses of the galaxies, as probed by observations of the 
various transitions of the CO molecule. We also compare and discuss the global
properties of the ``warm'' gas in Starbursts and Seyferts.

\subsection{Excitation Temperatures and Masses}

The excitation diagrams in Figures 3 and 4 show a plot of the
natural logarithm of the column density divided by the statistical
weight in the upper level of each transition against the energy
level. The column density follows from the Boltzmann equation,
\begin{equation}
\frac{N_{i}}{N} = \frac{g(i)}{Z(T_{ex})} \times exp(-\frac{T_{i}}{T_{ex}})
\end{equation}
where N$_{i}$ is the column density of H$_{2}$ in the i-th state,
g(i) and T$_{i}$ are the statistical weight and energy level of that
state, N is the total column density of H$_{2}$, T$_{ex}$ is the
excitation temperature, and Z(T$_{ex}$) is the partition function at
T$_{ex}$. N$_{i}$ is calculated from the observed quantities according
to:
\begin{equation}
N_{i} = \frac{flux(i)}{A(i)\times h \times \nu(i)} \times \frac{4\pi}{beam}
\end{equation}
where flux(i) is the flux of a line in the i-th state, A(i) is the
A-coefficient of that transition, $\nu$(i) is the frequency of the
transition, h is Planck's constant, and beam is the solid angle of the
beam size.  In Figures 3 and 4 we plot the natural logarithm of
eq. (1).
\begin{figure*}
\includegraphics[width=18cm]{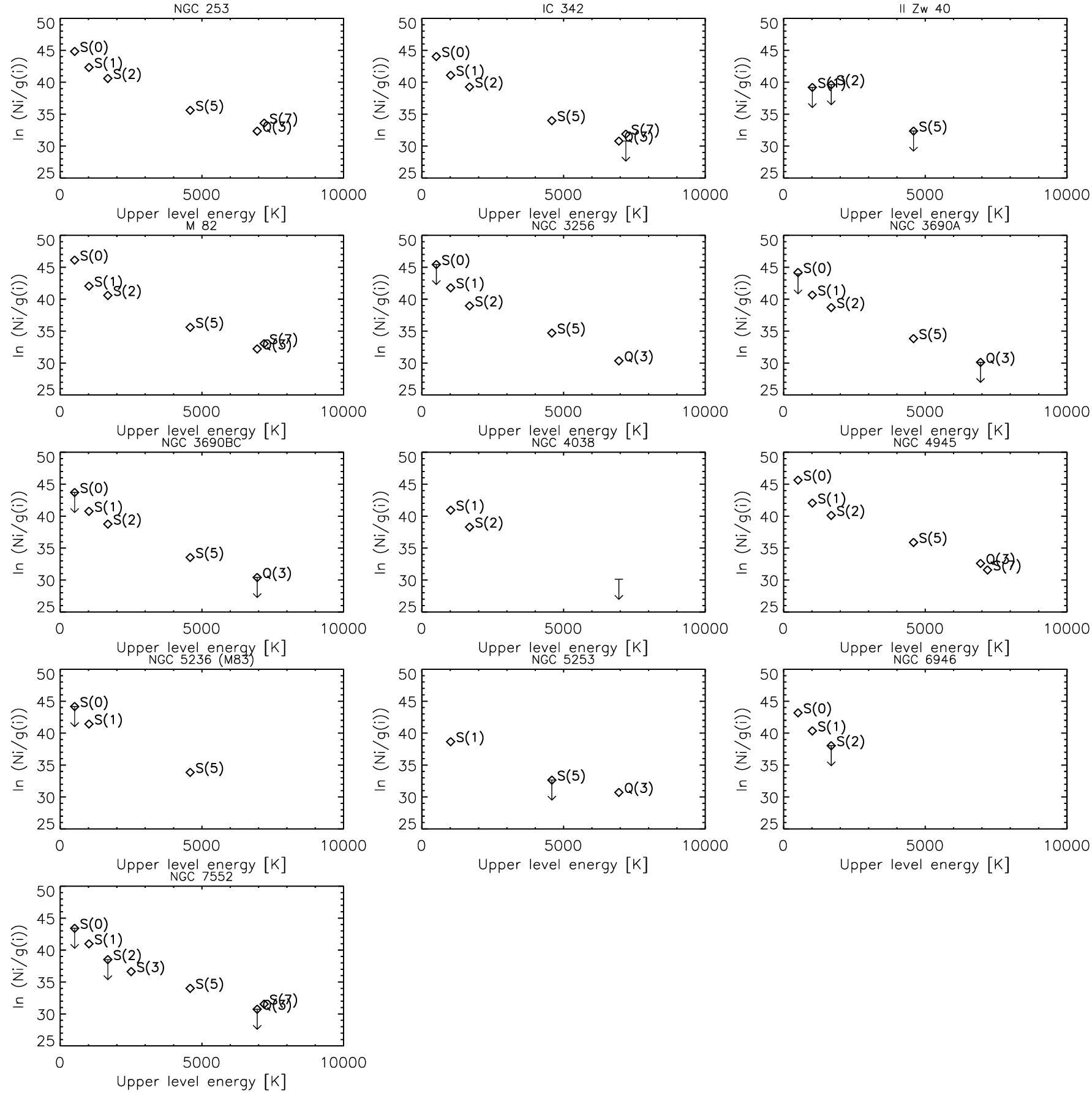}
\caption{Excitation Diagrams for the Starbursts. }
\label{fig3}
\end{figure*}

\begin{figure*}
\includegraphics[width=13cm,angle=90]{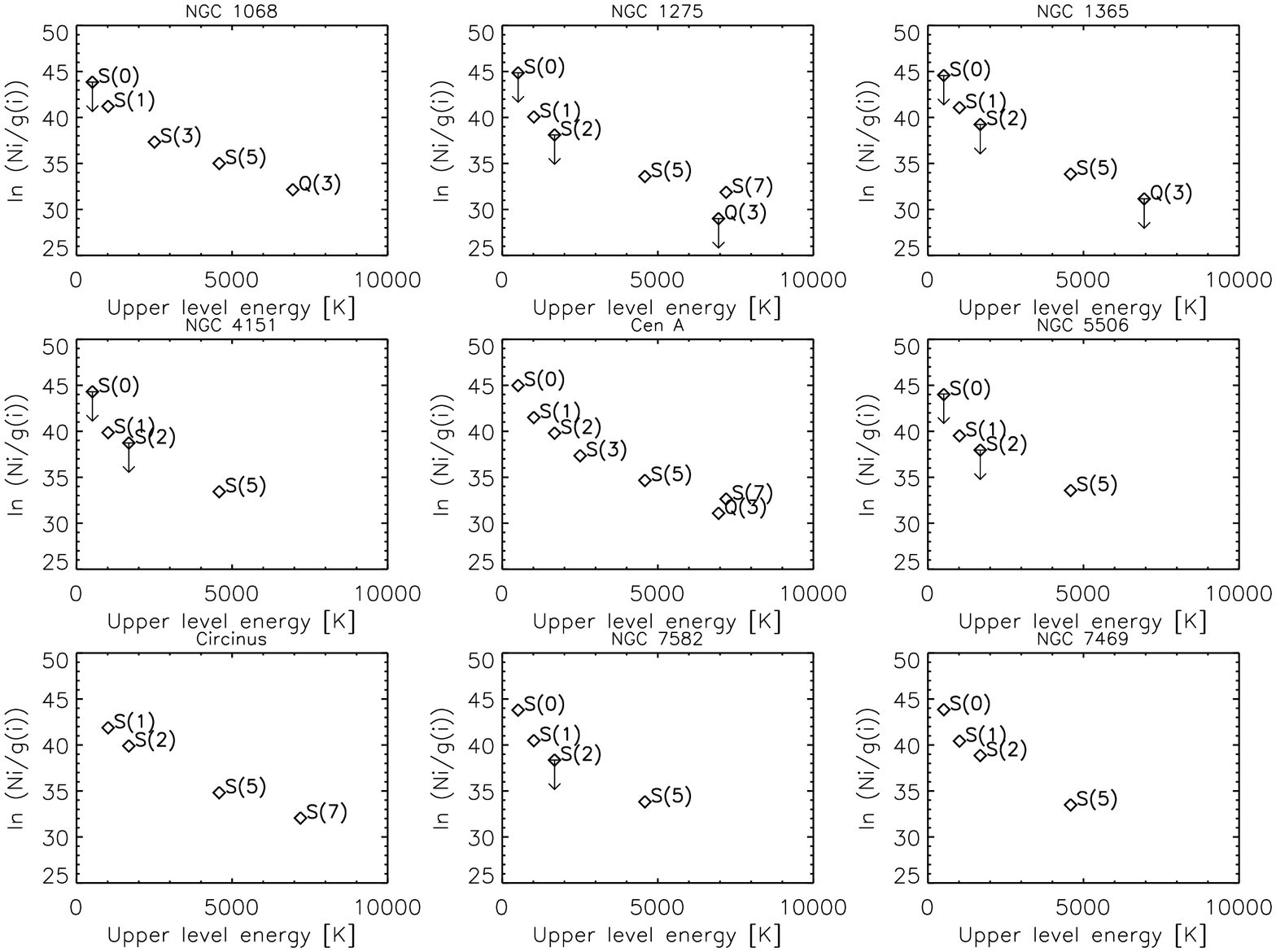}
\caption{Excitation Diagrams for the Seyferts.}
\label{fig4}
\end{figure*}

For the excitation diagrams we used extinction corrected line fluxes.
We estimated the extinction for each line flux using the A$_{V}$ 
values listed in Tables 1 and 2. For $\lambda<$8 $\mu$m we used 
A($\lambda$)$\propto \lambda^{(-1.75)}$ and for $\lambda>$8$\mu$m
the extinction law of Draine and Lee (1984) and Draine (1989) were used.
We assumed similar obscuration of the ionized and molecular media. 

The excitation temperature of the line emitting gas is the reciprocal
of the slope of the excitation diagram, corresponding to the kinetic
temperature in local thermodynamic equilibrium (LTE). Assuming
thermal emission, it is apparent that H$_{2}$ is present in a range of
temperatures in all cases\footnote{We note that the assumption of
LTE conditions implies densities higher than about n$>$10$^{4}$ cm$^{-3}$
which may not be present
if a diffuse component contributes significantly. In that case the
excitation diagram can still be used as an empirical tool although the
interpretation of excitation temperatures is not straightforward.}. The
excitation temperature changes rapidly with energy level. This is of
course a natural consequence of the fact that the gas is in reality
consisting of various components at various temperatures.  
The
detections of the S(1) and S(0) lines  (or the upper limits of the
latter whenever not detected) are used to constrain the temperature of
the ``warm'' line emitting  gas, while the S(5) and S(7) lines are
used to constrain the temperature of the somewhat ``hotter" gas. In
fact the S(5) and S(7) lines most likely probe  the same excited gas
as the near--infrared ro-vibrational lines.  Therefore, the detection of
the S(0) and S(1) lines is more important since we can probe the
more abundant ``warm'' gas.
In Tables 5 and 6 (columns 2 and 5 respectively) we list the 
temperatures (for the ``warm'' and the ``hotter'' gas)
for the Starbursts and  Seyferts, respectively.

We note that the derived excitation temperature is affected by the total 
ortho-to-para conversion rate of H$_{2}$. For the calculations above 
we have assumed an ortho-to-para abundance ratio of 3, the equilibrium 
value for temperatures T$\geq$200 K (assuming LTE conditions).  
Sternberg and Neufeld (1999) have measured an ortho-to-para ratio of 3 
in the PDR star-forming region S140. Assuming a lower ortho-to-para
conversion rate of 1 we find higher values of T$_{\small S(0)--S(1)}$
and lower values of T$_{\small S(2)--S(1)}$, internally inconsistent and
implying that multiple temperature components should be considered to fit 
the data. We believe that under the assumption of LTE conditions an ortho-to-
para abundance ratio of 3 is a realistic value.

From the S(0)--S(1) detections in Starbursts we derive an average 
temperature of T$_{\small SB} \sim$160$\pm$10 K. For those Starbursts 
for which there was no S(0) detection we used the S(1) and S(2) detections 
to find the range of temperatures. For the Starbursts we find 
150$\leq$T$_{\small SB}\leq$330 K.
For the Seyferts, the S(0) line is detected in CenA, Circinus, NGC 7582 and
NGC 7469. Based on these detections we derive an average temperature
T$_{\small Sey} \sim$ 150$\pm$12 K. 
For those Seyferts with S(0) non-detections we derive
limits in the range 120$\leq$T$_{\small Sey}\leq$370 K (using S(1) and S(2)).
Although the number of S(0) detections in the Seyfert sample is slightly
smaller
(in the present sample the S(0) line is detected in 5$/$10 of the 
Starbursts and 4$/$9 of the Seyferts) it is obvious
that the temperatures of the ``warm'' gas are, within the errors, 
similar in Starbursts and Seyferts.

Taken at face value the similar temperatures found for Starbursts
and  Seyferts indicate that the conditions in the gas are the same in
both  environments. Although this could have been the anticipated
result  (since we are looking at the same transitions originating in
``warm'' clouds  in galaxies) we caution that a number of factors may
contribute to this effect: First, most of the Seyferts in our
sample are ``mixed'' objects that  is, there is evidence for the
presence of an extended starburst component (e.g. NGC 7469, NGC 7582,
CenA and Circinus) whose energy output is comparable to that of the
central active nucleus.  Second, the large SWS apertures sample a
mixture of molecular 
hydrogen emission from cool clouds in the extended circumnuclear regions 
as well as the warmer central clouds. Thus, effects of 
dilution of a pure AGN-related effect by the more extended starburst activity
are possible. Such a behaviour has already been noted in the case 
of NGC 1068 (Lutz et al. 2000) where 
the SWS--H$_{2}$ emission traces gas originating both in the central active 
nucleus as well as the molecular ring which is present at larger spatial 
scales.

We next derive masses corresponding to the gas found at various 
temperatures, mostly the ``warm'' gas emitting the S(0) line as well as the
``hotter'' higher excitation gas in which the S(7) line originates. 
The various masses are derived based on the 
following method: the column density (from eq. 1) is multiplied by the physical
area of the object corresponding to the aperture. For aperture sizes we use the
values quoted in Section 2 depending on the transition (either S(0) or 
S(7)). 
The ``warm'' molecular masses were derived using the S(0) line 
(column density and temperature). In the cases of undetected S(0) lines  
we have used the S(1) line (column density) and the S(1)--S(2) 
temperature to derive the masses.
The ``hotter'' gas masses are derived using the S(7) detections. 
The ``warm'' gas masses (denoted as M$_{\small 1}$) and the ``hotter''
gas masses (denoted as M$_{\small 2}$) are listed in Tables 5 and 6 
(column 3 and 6) for Starbursts and AGN, respectively.

\section{Comparison with CO data}

Most of the knowledge about the molecular content of galaxies has so
far relied on observations of CO emission, assuming a
CO$/$H$_{2}$ conversion factor 
based on Galactic Giant Molecular Clouds (GMCs).
However, CO probes successfully gas 
lying at low temperatures of a few tens up to a
hundred. And although the low temperature gas comprises the bulk of
the gas content in a galaxy, it provides no information on the ``warmer''
gas which might be more directly linked to the source of activity.  In
section 3.2 we showed how ISO can give a firm estimate of the ``warm''
gas content of these galaxies.

Here, we try to estimate what fraction
of the total gas mass lies in higher temperatures as we directly
compare our new ISO findings with published CO data.
Although for such a comparison higher 
CO transitions (J= 3$\rightarrow$2, J=6$\rightarrow$5 etc) 
probing ``warmer'' gas (T $>$ 50K) are  more directly related to the gas 
seen by ISO however, these 
transitions are not available for the majority of the galaxies 
presented here. Such a detailed analysis has been carried out for only
two galaxies,
NGC 6946 (Valentijn et al. 1996), and NGC 4945 (Spoon et al. 2000). 

In Tables 5 and 6 we compare the masses derived from the SWS
H$_{2}$ emission lines with mass estimates from CO observations (assuming 
the GMC CO$/$H$_{2}$ conversion factor).
The comparison is more meaningful for galaxies where the S(0) is detected
because only then do we have an accurate estimate of the ``warm'' gas mass.

In the case of Starbursts the fraction of ``warm'' gas ranges between
1 to 10\% of the total gas mass. In Seyferts the fraction of the 
``warm'' gas is overall higher. Although there is great variation among the 
different objects, in general, the ``warm'' gas makes up between 
2 and 35\% of the total gas (as probed by CO), 
a value deduced by  using only those Seyferts with S(0) detections.
For the mass of the ``hotter'' gas 
we find that it accounts for a lot less, of the order of
at most 1\% in both Starbursts and Seyferts.

Although the gas in Starbursts and Seyferts has similar temperatures the 
corresponding masses of ``warm'' gas are distinctively different. The amount 
of ``warm'' gas in Seyferts is one to two orders of magnitude higher than in 
Starbursts. The origin of this difference could lie in the 
excitation mechanism. As we discuss in Section 6 apart from 
shocks and PDRs H$_{2}$ may also arise in gas heated by hard X-ray photons
originating in the central AGN. These X-ray photons are capable of penetrating
large columns of gas thus heating extensive parts of the circumnuclear regions
reaching out to the larger scale disks in AGN.  
Spatially resolved observations of rotational lines in AGN would be 
crucial in investigating the extent of the H$_{2}$ emitting regions.

\begin{table*}
\caption[]{\it Comparison of masses-Starbursts}
\begin{tabular}{cccccccc}
\hline
\hline
Galaxy Name&T(S(0)-S(1))&M$_{1}$&T(S(1)-S(2))&T(S(5)-S(7))&M$_{2}$&M$_{CO}$& 
Ref\\
  &K&$\times$10$^{8}$M$_{\odot}$&K&K&$\times$10$^{3}$M$_{\odot}$
&$\times$10$^{8}$M$_{\odot}$& \\
\hline
NGC 253&195&0.05 &380&1275&6.0&5&(1) \\
IC 342&171&0.05 &365&$\leq$1075&1.0&2&(1)  \\
IIZw40&-- &-- &--&--&-- &--& \\
M 82&120&0.54 &450&1020&4.0&6&(1) \\
NGC 3256&$\geq$135&10&230&--&-- &300&(1)\\
NGC 3690A&$\geq$140&-- &335 &-- &--  &130& (2)$^{a}$ \\
NGC3690B$/$C&$\geq$170&-- &-- &-- &-- &--&  \\
NGC 4038&--&-- &-- &-- &-- &-- &-- \\
NGC 4945&139&-- &-- &-- &-- &2.7&(3)\\
NGC5236 (M83)&$\geq$182&2.0 &-- &-- &--&--&-- \\
NGC5253&-- &-- &-- &--&--&-- \\
NGC6946&174&0.03 &$\leq$290&-- &-- &0.3& (4) \\
NGC7552&$\geq$158&0.8 &$\leq$272&1120&30&80&(1)\\
 & & & & & \\ \hline
\end{tabular}\\
References: (1) Aalto, et al., (1995); (2) Young et al. (1986);
(3) Henkel et al. (1994); (4) Sage et al. (1990); \\
$^{a}$: CO value refers to NGC 3690A$/$B
\end{table*}

\begin{table*}
\caption[]{\it Comparison of masses-Seyferts}
\begin{tabular}{ccccccccc}
\hline
\hline
Galaxy Name&T(S(0)-S(1))&M$_{1}$&T(S(1)-S(2))&T(S(5)-S(7))&M$_{2}$&M$_{CO}$&
Ref\\
  &K&$\times$10$^{8}$M$_{\odot}$&K&K&$\times$10$^{3}$M$_{\odot}$
&$\times$10$^{8}$M$_{\odot}$&Ref\\
\hline
NGC1068&$\geq$145&1.0 &-- &-- &-- &40 & (1)\\
NGC1275&$\geq$115&25&340&1170&3.0&60& (2)\\ 
NGC1365&$\geq$145.0&1.7 &365&-- &-- &--& \\
NGC4151&$\geq$115&0.1&$\leq$590&-- &-- &0.3&(3)  \\
CenA&145&0.11 &390&1300&2.0 &1.0&(4) \\
NGC5506&$\geq$112&2.0 &$\leq$430&-- &-- &3&(3)\\
Circinus&199&0.8 &221&555&1.53&30& (5)\\
NGC7582&153&0.76 &$\leq$315&-- &-- &100&(6) \\
NGC 7469&145&54&425&--  &-- &150&(3) \\
 & & & & & \\ \hline
\end{tabular}\\
References: (1) Sage et al., (1990); (2) Bridges and Irwin 1998;
(3) Rigopoulou et al. (1997); (4) Wild, Eckart \& Wilkind (1997);
(5): Curran et al. (1998); (6) Maiolino et al. (1997)
\end{table*}

\section{H$_{2}$ Excitation Mechanisms}

The H$_{2}$ molecule can be excited via three distinct mechanisms:\\
(a) {\it UV fluorescence} where 
photons with $\lambda \geq$ 912\.A are initially absorbed by the
H$_{2}$ molecule (in its Lyman and Werner bands) and later
re-emitted resulting in the population
of the various vibration--rotation levels of the ground state
(e.g. Black and van Dishoeck 1987).\\ 
(b) {\it Shocks} where
high--velocity gas motions in a quiescent cloud can
heat, chemically alter and accelerate the ambient gas resulting in 
excitation of the H$_{2}$ molecule (e.g. Hollenbach and McKee 1989) and,\\
(c) {\it X--ray illumination} where hard X-ray photons
are capable of penetrating deeply into molecular clouds and heating 
large amounts of gas 
(e.g. Maloney 1996).\\

To distinguish among the three primary excitation mechanisms is far
from  trivial: high UV radiation flux as well as high-velocity gas
motions are both present in the nuclei of Seyfert galaxies and (to a
lesser extent) of Starbursts.  UV radiation can originate either from
the existing stellar population, from current star-forming processes,
or the accretion disk. On the other hand,  shock excitation can
originate from the central engine, an outflow due to the large number
of supernovae remnants or finally gas motions due to galaxy-galaxy
interactions. In addition, the low rotational
transitions arise from collisionally excited warm gas which complicates the
distinction even further.

Since the signatures of all three mechanisms on the pure 
rotational lines are almost indistinguishable, 
an analysis including both ro--vribational and
rotational lines would be ideal to address the issue of the excitation
mechanism in extragalactic environments. However, the different apertures used
to measure lines at different parts of the spectrum
introduce large uncertainties, making such a study almost impossible.
In what follows we present a brief account of the various theoretical
model predictions and compare them to our observed line ratios. In addition,
we compare our measurements for the two extragalactic samples 
to well-studied Galactic templates with the aim of better understanding
the origin of the H$_{2}$ emission.

\subsection{Comparison with theoretical work}

\subsubsection{PDR models}

In this section we compare the observed line ratios to the published 
PDR models of Burton Hollenbach and Tielens (1992, BHT).
From the present samples we find mean line ratios of S(1)/S(0)=5.31$\pm$0.64
and 3.75$\pm$1.38 and S(1)/S(2)=2.33$\pm$0.82 and 1.85$\pm$0.93 for 
Starbursts and Seyferts, respectively. 
We find that the line ratios for the Starbursts can be explained with a 
sensible combination of gas densities and radiation field strengths. For 
instance the mean S(1)$/$S(0) Starburst value can be reproduced
by models 4 or 5 (from BHT) whereas to explain the S(1)$/$S(2) 
ratio a harder radiation 
field is necessary (such as in model 7). The mean S(1)$/$S(0) line ratio 
for Seyferts is somewhat but not significantly lower than the one in 
Starbursts. 
Model 3 reproduces fairly 
closely the S(1)$/$S(0) ratio in Seyferts but fail to predict the S(1)$/$S(2)
ratio which would again require a harder radiation field. 

Although it is impossible to constrain PDR models based on the current 
observations such direct comparisons of the model prediction and the current
measurements suggest that a combination of normal PDRs can explain the 
emission in both Starburst and Seyfert galaxies.

\subsubsection{Shock models}

A direct comparison of the various model predictions (e.g. Brand et al. 1989, 
Hollenbach \& McKee (1989), Kaufman \& Neufeld (1996)) with our observed
line ratios for Starbursts and Seyferts shows that, a single speed shock 
model cannot explain the observed line ratios. 
Additionally, almost all models predict S(1)$/$S(0) line ratios much larger 
than the observed ones (ie the predicted 0-0 S(0) emission is weaker).
If the gas temperature is high then the J=2 level is not populated resulting
in a weak 0-0 S(0) emission. In galactic sources with shocked H$_{2}$ 
emission, the temperature of the higher levels (say J=8 to J=3) is relatively 
constant at about 700 K (Wright, priv. comm.) thus, the J=2 level 
does not emit strongly. However, in extragalactic systems the situation 
is probably different. The low velocity shocks cannot heat the gas to 
high enough temperatures, the J=2 level is populated resulting in an
enhanced the 0-0 S(0) line emission compared to the various model 
predictions.  

Among the Seyferts, NGC 1275 is a candidate for shock induced H$_{2}$
emission. This is also supported by the ro-vibrational lines observed 
in the near-IR (Fischer et al. 1987, Krabbe et al. 2000).
For the Starburst galaxies it is harder to make any predictions:
Except perhaps from the
case of the luminous IR galaxy NGC 6240 (Egami et al. 2002, in preparation), 
no other
starburst galaxy is a clear case candidate for shock excited H$_{2}$
emission.

\subsubsection {X--ray irradiated molecular gas}

So far, the efforts to model X-ray Dominated Regions (XDRs) have been 
concentrated mostly on X--rays originating from a central engine. 
The model predictions account mostly for the 
ro-vibrational line ratios and$/$or consider additional diagnostics such
as the near-infrared [Fe II] 1.26 $\mu$m and the cooling lines [OI] 63 $\mu$m
and [C II] 158 $\mu$m (Maloney, Holenbach \& Tielens 1996).
Tin\'e et al. (1997) presented model predictions for the 
strength of the S(7) line, the {\em only} rotational transition 
so far included in XDR models. 
In accordance with the results of Maloney Holenbach \& Tielens (1996), Tin\'e 
et al. found that the H$_{2}$ emission arises
in typical temperatures T$<$1000 K and relatively high densities. For higher 
temperatures collisional excitation dominates the low vibrational levels. 
We note that the H$^{+}_{3}$ lines characteristic of X--ray irradiated gas
(Draine and Woods 1990) have not been detected in any of our sample Starbursts
or Seyferts, therefore it is difficult to make any predictions. An exception 
is NGC 1068 where upper limits in H$^{+}_{3}$ have been reported in 
Lutz et al. (2000) but, unfortunately, provide no constraints on the models.

From the comparisons of our measurements to theoretical models we conclude 
that much of the H$_{2}$ emission in 
Starbursts could originate in PDRs although a 
combination of PDRs and shocks of various speeds is also a viable solution. 
A similar situation is probably true for the Seyferts as well, although 
as we already discussed in Section 4  
X-rays from the central AGN may also play an important role in heating 
up large amounts of gas and triggering H$_{2}$ emission. 

Through the exercise of comparing our observations with theoretical models
we want to emphasize how difficult it is to discriminate between the various
H$_{2}$ emission mechanisms and that 
in order to investigate thoroughly the origin of the H$_{2}$ 
emission in galaxies a host of vibrational and rotational transitions 
must be available.

\subsection {The empirical Way: Comparison with Galactic Templates}

In this section we compare our measurements to a number of well studied
local Galactic templates.
For such a comparison we have considered two typical PDR regions ``S140'' 
(Timmermann et al. 1996) and the Orion Bar (Rosenthal 2000, priv. comm),
and two cases of shock-powered emission ``Cepheus A'' 
(Wright et al. 1996) and Orion Peak 1 (Rosenthal et al. 1999).
In Table 9 we list the temperatures and masses of the local Galactic templates
along with ``mean'' properties of our sample Starburst and Seyfert 
galaxies.  We note that for the mean values quoted in Table 9 we 
have used {\em all} sample galaxies. 

\begin{table*}
\caption[]{\it Comparison with Galactic Templates}
\begin{tabular}{cccc}
\hline
\hline
Type&S(1)$/$S(0)&S(1)$/$S(2)&T(S(0)--S(1))\\
 & & &K\\
\hline
Starbursts&5.3&2.33&160--(330)$^{1}$\\
Seyferts&3.75&1.85&150--(370)$^{1}$\\
S140 (PDR)&--&1.00&159\\
Orion Bar (PDR)&1.11&5.29&155 \\
Cepheus A West (shock)&  &0.54&700 \\
Orion Peak 1 (shock)&--&0.96&(570)$^{2}$\\
 & & &   \\ \hline
\end{tabular}\\
$^{1}$: The upper limit of 370 K is derived from S(1) and S(2) transitions
for those Seyferts where the S(0) is not detected.\\
$^{2}$: temperature value derived from S(1) and S(2) transitions.
\end{table*}

The average temperatures for the Starburst and Seyfert galaxies are the 
same as the temperature observed
in both cases of PDR emission and well below what is observed for 
an Orion type shock. Overall, although in reality different regions and 
excitation mechanisms 
contribute to the galaxy-integrated temperature distribution these simple 
comparisons suggest that normal PDRs are responsible for most of the emission 
in Starburst and Seyfert galaxies.

\section{PAH and H$_{2}$ emission}

It has been known for some time (see e.g. review by Puget \& L\`{e}ger 1989)
that the 3-12 $\mu$m part of the spectrum of galaxies is dominated by
a number of emission features that appear whenever the interstellar
medium  is exposed to moderately intense UV radiation. The exact
nature of the carriers that are responsible for the emission is most
likely large carbon-rich  molecules the so-called Polycyclic Aromatic
Hydrocarbons (PAH). With ISO it has been possible to detect PAH emission
in a number of galactic (e.g.  Verstraete et al. 1996, Roelfsema et
al. 1996) and extragalactic (e.g. Rigopoulou et al. 1999)
environments. 
Here, we want to investigate whether 
any link exists between PAH emission (ultimately related to the presence
of PDRs and HII regions) and rotational H$_{2}$ emission originating in 
the molecular clouds of galaxies where stars are formed. 
If both warm H$_2$ emission and PAH emission originate mostly in 
UV-illuminated surfaces of molecular clouds (i.e. PDRs), their luminosities 
should correlate well when averaged over a galaxy.

In Figure 5 we plot the 7.7$\mu$m PAH luminosity L$_{\small 7.7}$ 
vs. the H$_{2}$S(1) luminosity L$_{H_{2}S(1)}$ 
for our sample Starbursts and AGN.
\begin{figure}
\resizebox{\hsize}{!}{\includegraphics{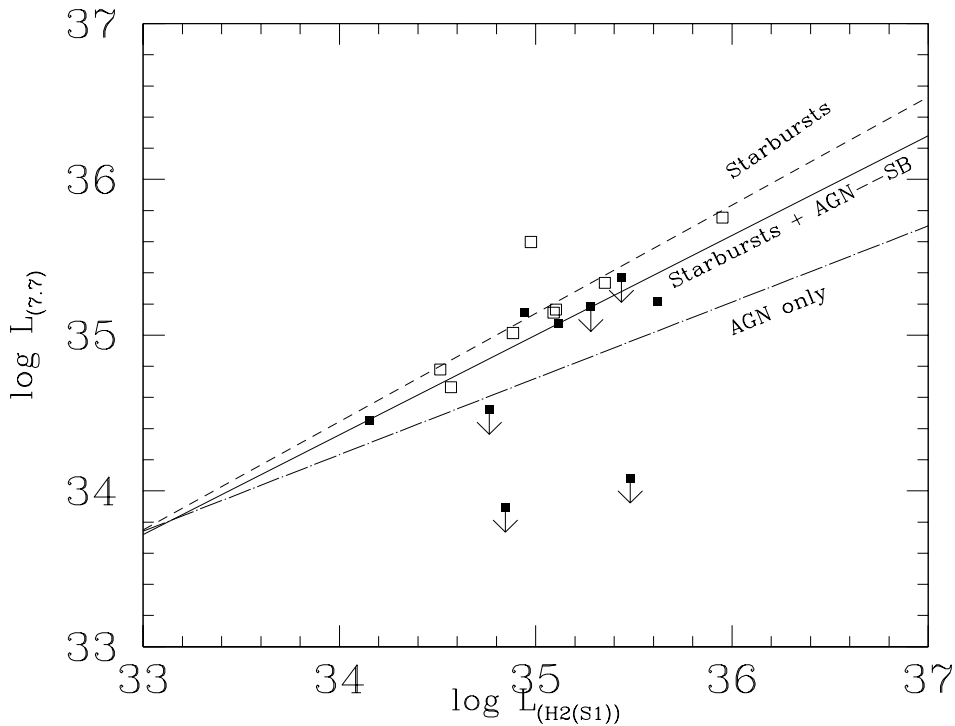}}
\caption{PAH luminosity L$_{\small 7.7}$ vs. the H$_{2}$S(1)
luminosity for our sample Starbursts and AGN. The open and filled
squares represent the Starburst and AGN points, respectively.
The lines shown represent the least square fits for :
dashed line: Starbursts, straight line: Starbursts and AGN detections, 
long--dashed line AGN only (non-detection included).}
\label{fig5}
\end{figure}
We find that Starbursts are consistent with a relation of the type 
log L$_{\small 7.7}$ =C$+$ 0.69 log L$_{\small H_{2}S(1)}$.
Some Seyferts deviate strongly from the H$_{2}/$PAH relation in having 
very weak
PAH for their H$_{2}$ luminosity. 
Seyferts as an entire sample (ie including those with PAH upper limits)
follow the relation  
log L$_{\small 7.7}$ =C$+$0.48 log L$_{\small H_{2}S(1)}$ which becomes
log L$_{\small 7.7}$ =C$+$0.53 log L$_{\small H_{2}S(1)}$ if we only use
AGN with firm PAH detections. 
Finally, we have calculated
the relation for both Starbursts and AGN (with firm detections) and found 
it to be log L$_{\small 7.7}$ = C $+$ 0.64log L$_{\small H_{2}S(1)}$, quite 
similar in slope to the one found for Starbursts.

Overall we find that the 7.7 $\mu$m and H$_{2}$S(1) luminosities of
Starbursts and AGN (with firm detections) are well correlated implying 
that the PAH and H$_{2}$ emission have a common starburst origin. 
Indeed, the four AGN with firm PAH detections, CenA, NGC 7582, 
Circinus and NGC 7469, all have substantial starburst activity (e.g.
Genzel et al. 1995 on NGC 7469, Maiolino et al. 1998 on Circinus etc).
Among the remaining five Seyferts three of them, NGC 4151, NGC 1275 and 
NGC 5506 are AGN dominated (with little or no Starburst contribution).
These three AGN are significantly
underluminous in PAH although they are relatively strong H2 emitters. 
A possible scenario to explain this behaviour is the following:\\
in Starbursts and in AGN with a significant large scale starburst component,
PAH and H$_{2}$ emission correlate well. In both cases the emission 
most likely originates in the Starburst. \\
In the remaining ``pure'' type AGN, H$_{2}$ emission is considerably 
stronger in comparison to the relatively weaker PAH emission. 
There are two possibilities for the high H$_{2}/$PAH ratio: 
either the PAH molecules are destroyed in the 
intense radiation field (possibly by the nuclear X-rays) or 
energetic photons
escaping from the central X-ray source heat a large fraction of the 
circumnuclear material which in turns gives rise to the 
enhanced H$_{2}$ emission (the PAH emission remains normal). 
Although it is very difficult to discriminate between these two
options the current observational evidence tends to favor the latter
possibility.
As we already discussed in Section 4 the gas masses responsible
for the H$_{2}$ emission comprise a larger fraction of the total available
gass mass (as traced by CO) in AGN than in Starbursts.

We conclude that the central AGN is probably responsible for the bi-modal 
behaviour of the PAH--H$_{2}$ relation in Starburst and Seyferts not 
by suppressing the PAH emission instead, by enhancing the H$_{2}$ emission 
by heating larger fractions of the available circumnuclear gas.

\section{Conclusions}

We have presented pure rotational H$_{2}$ emission lines observed with the 
ISO satellite from a sample of Starburst and Seyfert galaxies.
We have compared the emission properties of the two samples. The results are 
summarized as follows:\\

The lowest S(0) transition has been detected in 5 out of the 10  
Starbursts and 4 out of the 9 Seyfert galaxies. The S(1)$/$S(0) line ratio
is within the errors not that different in Starbursts and in Seyferts. 

The temperature of the ``warm'' gas, as estimated using the S(0) detections
was found to be similar (within errors) 
T$\sim$150 K, in both Starburst and Seyfert 
galaxies. This in turn implies that the global properties of the gas in both
environments are the same. We caution though that due to the large ISO beams
dilution of a pure AGN-related effect is important.\\

The ``warm'' gas mass constitutes up to 10\% of the total molecular gas 
content (as traced by CO molecular observations) in Starbursts.
However, the fraction of ``warm'' gas in Seyferts is considerably higher 
reaching up to 35\% (although with a large scatter in values). We propose
that, extra ``heating'' of the gas is provided by energetic hard X-ray photons
originating from the central AGN.

We have compared the observed strength of the molecular H$_{2}$ lines
with theoretical model predictions as well as those of local Galactic
templates.  Such comparisons reveal that a combination of various PDR
clouds explain  reasonably the line ratios observed in
Starbursts. Although such a combination of PDR models can also match
the observed line ratios in Seyferts it is likely that slow velocity
shocks and some heating from the central X-rays are also present.

Finally, we have examined the existence of a link between PAH and molecular 
line emission. We find that Starbursts and Seyferts with a strong
starburst-component follow a very similar correlation. On the other hand
the ratio H$_{2}/$PAH is higher in pure ``AGN--dominated'' objects. 
It is likely that an extended circumnuclear component of ``warm'' gas
(heated by the nuclear X-ray emission) is present in AGN 
in which enhanced H$_{2}$ emission originates. This ``warm'' gas component
is either too far away for UV photons to reach or, acts as a shield to UV 
photons resulting in both cases in suppressed PAH emission.

\acknowledgements{We thank Amiel Sternberg, Alfred Krabbe and the anonymous
referee for useful comments which significantly improved the present work. 
We thank Eckhard Sturm for help with the data reduction.
SWS and the ISO Spectrometer
Data Center at MPE are supported by DLR (DARA) under grants 50 QI 8610 8 and
50 QI 9402 3.}

{ }

\begin{thebibliography}{ }

\bibitem{} Aalto, S., Booth, R.S., Black, J.H., Johansson, L.E.B., 1995, 
\aap, 300, 369

\bibitem{}Alexander, T., Sturm, E., Lutz, D., et al., 1999, \apj, 512, 204

\bibitem{} Baldwin, J., Spinrad, H., Terlevich, R., 1982, \mnras, 198, 535

\bibitem{} Black, J.H., Dishoeck, E.F., 1987, \apj, 322, 412

\bibitem{} Bridges, T.J., Irwin, J.A., 1998, \mnras, 300, 967

\bibitem{} Burton, M.G., Hollenbach, D.J., Tielens, A.G.G., 1992, 
 \apj, 399, 563

\bibitem{} Curran, S.J., Johansson, L.E.B., Rydbeck, G., Booth, R.S., 1998, 
\aap, 338, 863

\bibitem{} deGraauw, Th., Haser, L.N., Beintema, D.A., et al., 1996, \aap,
315, L49

\bibitem{} Draine, B.T., 1980, \apj, 241, 1021

\bibitem{} Draine, B.T., 1989, in ESA, Infrared Spectroscopy in Astronomy, 93

\bibitem{} Draine, B.T., Roberge, W.G., Dalgarno, A., 1983, \apj, 264, 485

\bibitem{} Draine, B.T., Lee, H.M., 1984, \apj, 285, 89

\bibitem{} Draine, B.T., Woods, D.T., 1990, \apj, 363, 464

\bibitem{}Feuchtgr\"{u}ber, H., Lutz, D., Beintema, D.A., et al. 1997,
 \apj, 487, 962

\bibitem{}Fischer, J., Smith, H.A., Geballe, T. R.,
 Simon, M., Storey, J.W.V., 1987, \apj, 320, 667

\bibitem{} Genzel, R., Weitzel, L., Tacconi-Garman, L. E., et al., 1995, 
\apj, 444, 129

\bibitem{}Genzel, R., Lutz, D., Sturm, E., et al., 1998, \apj, 498, 579

\bibitem{} Habing, H.J., 1968, Bulletin of the Astronomical Institute of 
the Netherlands, vol. 19, 421

\bibitem{} Heckman, T.M., Butcher, H.R., Miley, G.K., 
 van Breugel, W.J.M., 1981, \apj, 247, 403

\bibitem{} Henkel, C., Whiteoak, J.B., Mauersberger, R., 1994, \aap, 284, 17
 
\bibitem{} Hollenbach, \& D., McKee, C.F., 1989, \apj, 342, 306

\bibitem{} Hyman, S.D., Lacey, C.K., Weiler, K.W., Van Dyk, S.D., 2000, 
\aj, 119, 1711

\bibitem{}Joseph, R.D., Wade, R., Wright, G.S., 1984, Nature, 311, 132

\bibitem{} Kaufman, M.J., Neufeld, D.A., 1996, \apj, 456, 611

\bibitem{}Kessler, M., Steinz,J.A., Anderegg, M.E., et al., 
 1996, \aap, 315, L27

\bibitem{}Krabbe, A., Sams, B.J., III, Genzel, R., Thatte, N., 
 Prada, F., et al. 2000, \aap, 354, 439

\bibitem{} Kristen, H.,  Jorsater, S., Lindblad, P., Boksenberg, A., 1997,
\aap, 328, 483

\bibitem{}Kunze, D., Rigopoulou, D., Lutz, D., et al. 1996, \aap, 315, L101

\bibitem{} Lutz, D., Genzel, R., Sturm, E., et al., 1997, ISO Spectroscopy of 
Luminous Galaxies, in Proceedings of the first ISO workshop on Analytical 
Spectroscopy, Heras, A.M.,Leech, K., Trams, N.R., Perry M., (eds), ESA Publications Division vol. 419, 143

\bibitem{}Lutz, D., Genzel, R., Sturm, E., et al. 2000, \apj, 536, 697

\bibitem{} Maloney, P.R., Hollenbach, D.J., Tielens, A.G.G.M., 1996, 
\apj, 466, 561

\bibitem{} Maiolino, R., Ruiz, M., Rieke, G.H., Papadopoulos, P., 1997, 
\apj, 485, 552

\bibitem{} Maiolino, R., Krabbe, A., Thatte, N., Genzel, R., 1998, 
\apj, 493, 650

\bibitem{} Maloney, P.R., 1996, \apss, 248, 105

\bibitem{} Moorwood, A.F.M., \& Oliva, E., 1988, \aap 203, 278

\bibitem{}Moorwood, A.F.M., Lutz, D., Oliva, E., et al., 1996, \aap, 315, L109

\bibitem{} Oke, J.B., Sargent, W.L.W., 1968, \aj, 73, 895

\bibitem{} Osterbrock, D.E., 1979, \aj, 84, 901

\bibitem{} Pug/'et, J.-L., \& L/`eger, A., 1989, \araa, 27, 161

\bibitem{}Puxley, P.J., Hawarden,T.G., Mountain, C.M., 1988, \mnras, 234, 29, 

\bibitem{}Rigopoulou, D., Lutz, D., Genzel, R., et al., 1996, \aap, 
 315, L125 

\bibitem{} Rigopoulou, D., Papadakis, I., Lawrence, A., Ward, M., 1997, 
\aap, 327, 493 

\bibitem{}Rigopoulou, D., Spoon, H.W.W., Genzel, R., et al., 1999, \aj, 
 118, 2625

\bibitem{} Roelfsema, P.R., Cox, P., Tielens, A.G.G.M., et al., 1996, 
\aap, 315, L289

\bibitem{} Rosenthal, D., Bertoldi, F., Drapatz, S., 1999, \aap, 356, 705

\bibitem{} Sage, L.J., Shore, S.N., Solomon, P.M., 1990, \apj, 351, 422

\bibitem{} Sanders, D.B., Scoville, N.Z., Soifer, B.T., 1991, \apj, 370, 158

\bibitem{}Schaeidt, S.G., Morris, P.W., Salama, A., et al. 1996, 
 \aap, 315, L55

\bibitem{} Spoon, H.W.W., Koornneed, J., Moorwood, A.F.M., et al., 2000,
\aap, 357, 898

\bibitem{} Sternberg, A., \$ Neufeld, D.A., 1999, \apj, 516, 371

\bibitem{} Timmermann, R., Bertoldi, F., Wright, C.M., et al., 1996, 
 \aap, 315, 281

\bibitem{} Tin\'e, S., Lepp, S., Gredel, R., Dalgarno, A., 1997, \apj 481, 282

\bibitem{} Usuda, T., Sugai, H., Kawabata, H., et al., 1996, \apj, 464, 818

\bibitem{}Valentijn, E.A., van der Werf, P.P., de Graauw, T., 
 deJong, T., 1996, \aap, 315, 145

\bibitem{} Valentijn, E.A., van der Werf, P.P., 1999, \apj 522, 29

\bibitem{} Van der Werf, P.P., Genzel, R., Krabbe, A., et al., 1993, 
 \apj, 405, 522

\bibitem{} Verstraete, L., Puget, J.-L., Falgarone, E., et al., 1996, \aap,
 315, L337

\bibitem{} Wild, W., Eckart, A., Wiklind, T., 1997, \aap, 322, 419

\bibitem{} Wright, C.M., Drapatz, S., Timmermann, R., et al. 1996, 
\aap, 315, L301

\bibitem{} Young, J.S., Kenney, J.D., Tacconi, L.J., et al., 1986, 
\apj, 311, L17

\end{thebibliography}
\end{document}